\documentclass[a4paper]{scrartcl}
\usepackage{epsfig}
\usepackage{amssymb}
\usepackage{graphicx,color}
\usepackage{amsmath}
\usepackage{array}
\usepackage[center]{subfigure}
\usepackage{slashed}
\usepackage{bbm}
\usepackage{mathrsfs}
\usepackage{ulem}
\usepackage{multirow}
\usepackage{hyperref}
\newcommand{\bra}[1]{\left\langle #1 \right|}
\newcommand{\ket}[1]{\left| #1 \right\rangle}

\newcommand{\be}{\begin{equation}}
\newcommand{\ee}{\end{equation}}
\newcommand{\ba}{\begin{eqnarray}}
\newcommand{\ea}{\end{eqnarray}}

\begin{document}

 \begin{titlepage}
\pagestyle{empty}
\begin{flushright}
SI-HEP-2022-08\\
P3H-22-060
\end{flushright}
\vfill
 \begin{center}
 {\Large \boldmath
 \textbf{$B$-meson decay into a proton 
and dark antibaryon from QCD light-cone sum rules }\\[.5cm]}
 
{\large Alexander Khodjamirian  and Marcel Wald }
\vspace*{0.4cm}

\textsl{
Center for Particle Physics Siegen (CPPS), Theoretische Physik 1\,,\\
Universit\"at Siegen, D-57068 Siegen, Germany \\[3mm]
}

\begin{abstract}
 The recently developed $B$-Mesogenesis scenario
 predicts decays of $B$ mesons into a baryon and hypothetical dark antibaryon $\Psi$. 
 We suggest a  method 
 to calculate the amplitude  of the simplest exclusive decay mode $B^+\to p \Psi$. 
 Considering two models of $B$-Mesogenesis, we obtain the $B\to p$ hadronic matrix elements by applying QCD light-cone sum rules  with the proton light-cone distribution 
 amplitudes. We estimate the $B^+\to p \Psi$
 decay width as a function of the mass and effective coupling of the dark antibaryon. 
 
\end{abstract}
\end{center}

 \vfill
 \end{titlepage}

{\large \bfseries 1.}~~
The $B$-Mesogenesis scenario  suggested in \cite{Elor:2018twp,Alonso-Alvarez:2021qfd} (see also \cite{Elahi:2021jia,Alonso-Alvarez:2021oaj}) simultaneously solves the problems of
 the baryon asymmetry and relic dark matter abundance in the Universe. 
The effective four-fermion interactions emerging  in the 
$B$-Mesogenesis models  generate decays of the $b\,$-quark into a  light diquark and a dark antibaryon $\Psi$ such that the baryon charge is conserved.
 These processes trigger new exotic decay modes  
of $B$ mesons and $\Lambda_b$ baryons into final states containing light hadrons and $\Psi$, the latter revealing itself as an "invisible" i.e. a missing energy-momentum in the particle detector.

According to \cite{Elor:2018twp,Alonso-Alvarez:2021qfd}, the decays $B\to hadrons +\Psi$ should have 
appreciable branching fractions for a realization of the  $B$-Mesogenesis scenario, with an inclusive width in the ballpark of $10^{-4}$. It is therefore very important to estimate the branching fractions of separate exclusive decay modes in order to investigate if these modes are within the reach of the ongoing searches for $B$ meson decays to invisibles. In \cite{Alonso-Alvarez:2021qfd}, only the ratios of exclusive and inclusive widths were
estimated using a phase-space counting of quark states.
Here, we consider the simplest two-body decay $B^+\to p +\Psi$ with a single proton and missing energy in the final state. 
 Since a hadronic  $B\to p$ transition is involved, albeit an unusual one, we deal with the problem of calculating the hadronic matrix element of this transition within a QCD-based framework.

In this letter, we demonstrate how to solve this problem using the QCD light-cone sum rules.
This method introduced in
\cite{Balitsky:1986st,Balitsky:1989ry, Chernyak:1990ag} 
and applied to various  hadronic matrix elements
can easily be extended to the meson-baryon matrix elements. 
To this end, we introduce a proton-to-vacuum correlator of the $B$-meson interpolating current with the effective three-quark operator coupled to the $\Psi$ particle.
We calculate this correlator near the light-cone in terms of nucleon distribution amplitudes (DAs). After that we derive the LCSRs 
for the $B\to p$ form factors, employing dispersion relations and quark-hadron duality
in the $B$-meson current channel.
Earlier, the LCSRs with nucleon DAs have been used to calculate the nucleon electromagnetic form factors 
\cite{Braun:2001tj,Braun:2006hz,Anikin:2013aka}
and the semileptonic form factors of heavy baryons (see e.g.  \cite{Khodjamirian:2011jp}).
On the other hand, the $B$-meson interpolating current is widely used in the LCSRs for various 
$B\to h$ form factors, where $h$ is a light meson. Hence, there are practically no new 
or unknown inputs in the new sum rules obtained below.\\

{\large \bfseries 2.}
According to \cite{Elor:2018twp, Alonso-Alvarez:2021qfd}, the
decay of a $B^+$ meson into a proton and dark antibaryon $\Psi$ is generated by an exchange of a 
 heavy colour-triplet scalar field $Y$. This model has two versions  
 which differ by the electric charge $Q_Y$ of the heavy mediator. 
 In this paper, for definiteness, we consider only the version with $Q_Y=-1/3$, where
 the interactions of the $Y$-field with the  quarks and $\Psi$ are encoded in the 
 Lagrangian: 
  \ba
{\cal L}_{(-1/3)}=-y_{ud}\epsilon_{ijk}Y^{*\,i}\bar{u}_R^{\,j}d_R^{c\,k}
 -y_{ub}\epsilon_{ijk}Y^{*i}\bar{u}_R^{\,j}b_R^{c\,k}
 -y_{\Psi d}Y_i\bar{\Psi} d_R^{c\,i}-y_{\Psi b}Y_i\bar{\Psi} b_R^{c\,i}+ \mathrm{h.c.}\,,
 \label{eq:L}
 \ea 
 where the index $c$ ($R$) indicates  charge conjugated (right-handed) fields, 
 $q_R=\frac12(1+\gamma_5)q$, $i,j,k$ are the colour indices and 
 only  the terms relevant for the $B\to p\Psi$ decay  are shown. 
 
 The effective four-fermion interactions generating this decay 
  are given in \cite{Elor:2018twp, Alonso-Alvarez:2021qfd} in a generic form.
 Here we  derive more detailed expressions with a definite Dirac structure. The effective interactions 
 are obtained by forming the $Y$ exchange diagrams
 with the two vertices taken from the 
Lagrangian ${\cal L}_{(-1/3)}$  and neglecting the momentum transfer 
$\lesssim m_B$ with respect to a very large mass $m_Y$ in the propagator.
For the model (\ref{eq:L}), we obtain 
the local effective Hamiltonian:
\ba
 {\cal H}_{(-1/3)}\, =
 - \frac{y_{ub}y_{\Psi d}}{M_Y^2} i \epsilon_{ijk}\left(\bar{\Psi}d_R^{c\,i}\right)\left(\bar{u}_R^{j} b_R^{c\,k}\right)
 - \frac{y^*_{ub}y^*_{\Psi d}}{M_Y^2} i \epsilon_{ijk}\left(\bar{b}^{c\,i}_R u^{\,j}_R\right)\left(\bar{d}^{c\,k}_R\Psi\right)
 +
 \{d\leftrightarrow b \}\,.
 \label{eq:Heff1}
 \ea
We find it convenient to rewrite the above formula 
with the charge conjugation 
matrix attributed to the dark fermion field $\Psi$.
 To this end, we use the equations \footnote{Similar equations were derived in the context of generalized 
 Fierz transformations in \cite{Nieves:2003in}.} 
 \be
 \bar{\Psi}d_R^c=\bar{d}_R \Psi^c, ~~\bar{d}_R^c\Psi=\bar{\Psi}^cd_R \; ,
 \label{eq:transf}
 \ee
  which can be proven by employing  
 the explicit form of the 
 charge conjugation transformation for a Dirac field. 
 We then factorize out the "external" $\Psi$ field
 from the  effective Hamiltonian:
 \ba
 {\cal H}_{(-1/3)} \, =
 - \; G_{(d)}\bar{{\cal O}}_{(d)} \Psi^c 
  -\:
 G^*_{(d)}\bar{\Psi}^c{\cal O}_{(d)}+
 \{d\leftrightarrow b \}\,,~~~~
 \label{eq:Heff2}
 \ea
 where $G_{(d)}=(y_{ub}y_{\Psi d})/M_Y^2$
 is the effective four-fermion coupling 
 and 
 \be
 \bar{{\cal O}}_{(d)}= i \epsilon_{ijk}\left(\bar{u}^i_{R} b_R^{c\,j}\right)\bar{d}^k_R, ~~
 {\cal O}_{(d)}= i \epsilon_{ijk}d^{\,i}_R \left(\bar{b}^{c\,j}_R u^{\,k}_R\right) 
 \label{eq:Od}
 \ee
 are the local three-quark operator and its conjugate. The index $(d)$
 indicates  the terms in the effective Hamiltonian (\ref{eq:Heff2})
 originating from the $d$-quark coupling 
 to $\Psi Y$. Replacing $d\leftrightarrow b $ yields terms in (\ref{eq:Heff2}) 
 with the effective coupling $G_{(b)}=(y_{ud}y_{\Psi b})/M_Y^2$
and, respectively, with the three-quark operators
 \be
 \bar{{\cal O}}_{(b)}= i \epsilon_{ijk}\left(\bar{u}^i_{R} d_R^{\,c\,j}\right)\bar{b}^k_R, ~~
 {\cal O}_{(b)}= i \epsilon_{ijk}b^{\,i}_R \left(\bar{d}^{\,c\,j}_R u^{\,k}_R\right) \; .
 \label{eq:Ob}
 \ee
We treat the two parts of  ${\cal H}_{(-1/3)} $ with 
 the operators (\ref{eq:Od}) and (\ref{eq:Ob}) as two
 sample models of  the $B$-Mesogenesis, denoting them 
 as model $(d)$ and model $(b)$, respectively.
In \cite{Alonso-Alvarez:2021qfd} they correspond to the "type 2" and "type 1" 
 operators. \\

\noindent{\large \bfseries 3.}
We  consider  first the model $(d)$ in which the decay $B\to p \Psi$ is generated by the 
effective interaction involving the three-quark operator ${\cal  O}_{(d)}$.
It is then straightforward 
to write down the decay amplitude
\ba
{\cal A}_{(d)}(B^+\to p \Psi)&=& G_{(d)}
\langle p(P)| \bar{{\cal  O}}_{(d)} |B^+(P+q)\rangle  u^c_{\Psi}(q)\,,
\label{eq:ampld}
\ea
where $u^c_\Psi$ is the (charge conjugated) bispinor of the $\Psi$ field.
The four-momentum assignment is shown explicitly in \eqref{eq:ampld} and the on-shell conditions are $P^2=m_p^2$\,, $(P+q)^2=m_B^2$,
$q^2=m_\Psi^2$. 

The hadronic 
matrix element of the three-quark operator in (\ref{eq:ampld}) can be parameterized 
in terms of four independent $B\to p$ form factors: 
\ba
\langle p(P)| \bar{{\cal  O}}_{(d)} |B^+(P+q)\rangle =
F^{(d)}_{B\to p_R}(q^2)\bar{u}_{pR}(P)+
F^{(d)}_{B\to p_L}(q^2)\bar{u}_{pL}(P)
\nonumber\\
+
\widetilde{F}^{(d)}_{B\to p_R}(q^2)\bar{u}_{pR}(P) \frac{\slashed{q}}{m_p}+ \widetilde{F}^{(d)}_{B\to p_L}(q^2)\bar{u}_{pL}(P) \frac{\slashed{q}}{m_p}\,,
\label{eq:ff}
\ea 
where $u_{p\,R,L}= \frac12 (1 \pm \gamma_5)u_p$ is the bispinor of the right-handed (left-handed) proton.
Indeed, there are four independent kinematical structures
 constructed using the bispinor and two four-momenta $P$ and $q$.
Note that the structures $\bar{u}_{p\,L,R}(P)\slashed{P}$ are
reduced to $\bar{u}_{p\,L,R}(P)$ 
due to the Dirac equation.
   
Substituting (\ref{eq:ff}) into (\ref{eq:ampld}) at fixed $q^2=m_\psi^2$ and using the
Dirac equation for a charge-conjugated  bispinor, 
$~\slashed{q}\,u^c_{\Psi}(q)=m_\Psi\,u^c_{\Psi}(q)~$,
we obtain the decay amplitude in a compact form:
\ba
{\cal A}_{(d)}(B^+\to p\Psi)= 
G_{(d)}\bar{u}_{p}(P)\left(A^{(d)} +
B^{(d)}\gamma_5\right)u^c_{\Psi}(q)\,.
\label{eq:amplAB}
\ea
Here, the coefficients in the scalar and pseudoscalar structures are linear combinations 
of the $B\to p$ form factors: 
\ba
\begin{array}{c} A^{(d)}\\B ^{(d)}\end{array}\Bigg\}=
\pm\frac12\left(F^{(d)}_{B\to p_{R}}(m_{\Psi}^2)
+ \frac{m_\Psi}{m_p} \widetilde{F}^{(d)}_{B\to p_{R}}(m_{\Psi}^2)\right)
\nonumber\\
+\frac12\left(F^{(d)}_{B\to p_{L}}(m_{\Psi}^2)
+ \frac{m_\Psi}{m_p} \widetilde{F}^{(d)}_{B\to p_{L}}(m_{\Psi}^2)\right)\,.
\label{eq:ampl}
\ea

In order to square the amplitude (\ref{eq:amplAB})
and perform the sum over the proton and $\Psi$ polarizations, we use
the relation
\be
\overline{|\bar{u}_{p}(P)\Gamma u^c_{\Psi}(q) |^2}
\nonumber\\
=\mbox{Tr}\{(\slashed{P}+m_p)\Gamma
(\slashed{q}+m_\Psi)\gamma^0\Gamma^\dagger\gamma^0 \}
\label{eq:sq}
\ee
with $\Gamma=A^{(d)} + B^{(d)}\gamma_5$.
Multiplying the squared amplitude by the kinematical factors, 
we obtain the two-body decay width in terms of the 
form factor combinations defined in (\ref{eq:ampl}):
\ba
\Gamma_{(d)}(B^+\to p\Psi)\!\!\!&&=\,
|G_{(d)}|^2\,\Bigg\{|A^{(d)}|^2
\big(m_B^2-(m_p-m_\psi)^2\big) 
\nonumber\\
&&+\, |B ^{(d)}|^2\big(m_B^2-(m_p+m_\psi)^2\big)\Bigg\}
\,\frac{\lambda^{1/2}(m_B^2,m_p^2,m_\Psi^2)}{8\pi m_B^3}\,,
\label{eq:width}
\ea
where $\lambda$ represents the K\"allen function. 

Turning to the model $(b)$ with the operator ${\cal  O}_{(b)}$,
one has to replace $d\to b$ in the above equations (\ref{eq:ampld})-(\ref{eq:width}).\\


{\large \bfseries 4.}~~
Here we derive the LCSRs for the $B\to p$ form factors 
starting from the model with the three-quark  operator ${\cal O}_{(d)}$.
The key element of our method is 
the correlation function
\be
\Pi^{(d)}(P,q)=i\int d^4x\ e^{i(P+q)\cdot
x}\bra{0}T\left\{j_B(x),{\cal O}_{(d)} (0)\right\}\ket{p(P)} \,.
\label{eq:corr}
\ee
It contains the time-ordered product of 
${\cal O}_{(d)}$ and 
the $B$-meson interpolating current 
 $j_B=m_b\,\bar{b}i\gamma_5u$ with the four-momentum $P+q$. 
 The bilocal operator product in (\ref{eq:corr}) is 
sandwiched between the  on-shell proton state and the vacuum.
The correlation function
(\ref{eq:corr}) is decomposed in kinematical structures:
\ba
\Pi^{(d)}(P,q)=\Pi^{(d)}_R((P+q)^2,q^2) u_{pR}(P)+\Pi^{(d)}_L((P+q)^2,q^2) u_{pL}(P)
\nonumber\\
+\widetilde{\Pi}^{(d)}_R((P+q)^2,q^2)\slashed{q} u_{pR}(P)+
\widetilde{\Pi}_L^{(d)}((P+q)^2,q^2)\slashed{q} u_{pL}(P)\,,
\label{eq:corrdecomp}
\ea
where $\Pi_{R,L},\widetilde{\Pi}_{R,L}$ are Lorentz-invariant
amplitudes.

If the external momenta $P+q$ and $q$
 are far off-shell, so that 
$(P+q)^2\ll m_b^2$  and $q^2\ll m_b^2$, the integration region 
 in (\ref{eq:corr})  effectively shrinks to a domain near the
light-cone $x^2\sim 0 $, where the propagating $b$-quark is highly virtual. In this region, we 
calculate the correlation function
in terms of perturbative amplitudes convoluted 
with the light-cone distribution amplitudes (DAs) 
of the proton. 
Similar correlation functions were used to obtain the LCSRs for
$\Lambda_b\to p$ form factors \cite{Khodjamirian:2011jp}.

Below we will employ the definitions of the  nucleon three-quark DAs  in which the proton 
is in the initial state, hence our choice for 
the initial and final state in (\ref{eq:corr}). 
As a result, we will actually obtain LCSRs for  
the inverse hadronic transition $p \to B$. Accordingly, (\ref{eq:corr})
contains the three-quark operator 
${\cal O}_{(d)}$ from (\ref{eq:Od})
instead of its Dirac conjugate in the 
decay amplitude (\ref{eq:ampld}). This choice is in fact inessential, because 
the $p\to B$ form factors that will 
be obtained from LCSRs coincide with the $B\to p$ form factors in the decay amplitude
up to an irrelevant global phase. 

To access the form factors from the correlation function (\ref{eq:corr}),
we employ the hadronic dispersion relation in the variable $(P+q)^2$: 
\ba
\Pi^{(d)}(P,q)=\frac{\langle 0 | j_B|B^+(P+q)\rangle
\langle B^+(P+q)| {\cal  O}_{(d)} |p(P)\rangle
}{m_B^2-(P+q)^2}+
\int\limits_{s_h} ^\infty ds\frac{\rho^{h(d)}(s,P,q)}{s-(P+q)^2}\,,
\label{eq:hadrdisp}
\ea
where we isolate the $B$-meson pole. The contributions
of the excited and continuum states located above the lowest threshold $s_h=(m_B+2m_\pi)^2$ are collected
in the integral over  the spectral density $\rho^{h(d)}$. 
Note that subtractions are neglected in (\ref{eq:hadrdisp})
in anticipation of the Borel transformation.
Equation (\ref{eq:hadrdisp}) yields
dispersion relations for the invariant amplitudes defined in (\ref{eq:corrdecomp}).
To separate them, we rewrite the proton-to-$B$ hadronic matrix element in terms of 
form factors and use, analogously to (\ref{eq:corrdecomp}), the decomposition into invariant functions for the hadronic spectral density $\rho^{h(d)}$. We obtain for the first invariant amplitude

\ba
\Pi^{(d)}_R((P+q)^2,q^2)=\frac{m_B^2f_B\,F^{(d)}_{B\to p_R}(q^2)}{m_B^2-(P+q)^2}+ \;
\!\!\!\int\limits_{s_h} ^\infty\!ds\frac{\rho_R^{h(d)}(s,q^2)}{s-(P+q)^2}\,,
\label{eq:hadrdispR}
\ea
where we use the definition of the $B$-meson decay constant:
$\langle 0 | j_B|B^+\rangle=m_B^2f_B$. The remaining three dispersion relations are obtained 
from the above by replacing, respectively, 
\ba
&&
\Pi^{(d)}_R\to\Pi^{(d)}_L,~\widetilde{\Pi}^{(d)}_R,~\widetilde{\Pi}^{(d)}_L,  
~~~F^{(d)}_{B\to p_R} \to F^{(d)}_{B\to p_L}, ~m_p^{-1}\widetilde{F}^{(d)}_{B\to p_R},~
m_p^{-1} \widetilde{F}^{(d)}_{B\to p_L}\,,
\nonumber\\
&&
\rho_R^{h(d)}\to \rho_L^{h(d)},~ \widetilde{\rho}_R^{\,h(d)},~ \widetilde{\rho}_L^{\,h(d)}\,. 
\label{eq:repl}
\ea
 Note that in (\ref{eq:hadrdispR}) we tacitly replace the $p\to B$ 
form factors by the $B\to p$ ones,
ignoring an irrelevant global phase. 

Below, we will calculate
the correlation function
(\ref{eq:corr}), employing  the light-cone OPE in terms of the proton DAs. 
The results for invariant amplitudes will be cast in a  
convenient form of dispersion relations, e.g. for the amplitude $\Pi_R^{(d)}$:
\be
\Pi^{(d)OPE}_R((P+q)^2,q^2)=\frac{1}{\pi}\int\limits_{m_b^2} ^\infty ds\frac{\mbox{Im}\Pi_R^{(d)OPE}(s,q^2)}{s-(P+q)^2}\,.
\label{eq:OPEdsip}
\ee
This representation allows us to employ
the usual assumption of (semi-global) quark-hadron duality:

\be
\int\limits_{s_h} ^\infty ds\frac{\rho_R^{h(d)}(s,q^2)}{s-(P+q)^2}=
\frac{1}{\pi}\int\limits_{s_0^B} ^\infty ds\frac{\mbox{Im}\Pi_R^{(d)OPE}(s,q^2)}{s-(P+q)^2}
\,,
\label{eq:dual}
\ee
introducing the effective threshold  $s_0^B$.

Substituting (\ref{eq:OPEdsip}) 
into (\ref{eq:hadrdispR}), we then subtract from both parts of the resulting equation 
the integrals that are equal due to the duality approximation (\ref{eq:dual}). 
Performing the standard 
Borel transformation $(P+q)^2\to M^2$, we finally arrive at the LCSR for the first form factor:
\be
m_B^2f_B\,F^{(d)}_{B\to p_R}(q^2)e^{-m_B^2/M^2}=
\frac{1}{\pi}\int\limits_{m_b^2} ^{s_0^B} ds\, e^{-s/M^2}\mbox{Im}\Pi_R^{(d)OPE}(s,q^2)\,.
\label{eq:LCSR} 
\ee
The sum rules for the other three form factors are obtained from the above one by 
the replacements (\ref{eq:repl}). 

The main advantage of our method is its apparent universality. Turning to the model $(b)$,
 we derive the LCSRs for the corresponding form factors, 
repeating the same steps as above.
We only need to replace the three-quark operator 
${\cal O}_{(d)}$ by ${\cal O}_{(b)}$
in the 
correlation function (\ref{eq:corr}).\\

\begin{figure}
\begin{center}
\includegraphics[scale=1.3]{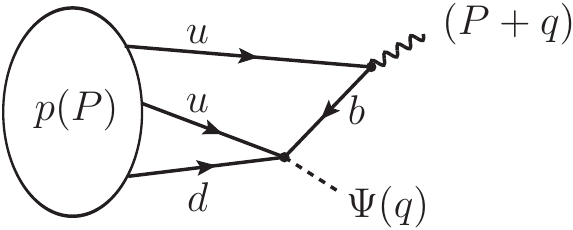}
\end{center}
\caption{Diagrammatic representation of the correlation function (\ref{eq:corr}). The wavy (dashed) line represents the interpolating current of the $B$ meson (the dark antibaryon).}
\label{fig:diag}
\end{figure}

{\bfseries 5.}~~To calculate the correlation function (\ref{eq:corr}), we 
write down the operators in terms of  quark fields:
\ba
&&
\Pi^{(d)}_\alpha(P,q)= \; - i \epsilon^{ijk} m_b\int d^4x\ e^{i(P+q)\cdot x}
\nonumber \\
&&\times
\bra{0}T\left\{\big[\bar{b}(x)\gamma_5u(x)\big] ,
d_R^i(0)_\alpha\big[\big(b^{j}_R(0)\big)^TCu_R^k(0)\big]\right\}\ket{p(P)} \,,
\label{eq:corrq}
\ea
so that the Dirac indices  inside the square brackets are contracted. The index $T$ denotes a transposed field, and $C$ is the charge 
conjugation matrix. 
We hereafter neglect the $u,d$-quark masses and assume isospin symmetry.

In the deep spacelike region, $(P + q)^2, q^2\ll m_b^2$, the expansion near the light-cone, $x^2=0$,  is applied to the operator product in (\ref{eq:corrq}).
To leading  $O(\alpha_s^0)$, the correlation function is described by the 
diagram shown in Fig.~\ref{fig:diag}, in which the perturbative part 
is reduced  to the free $b$-quark propagator and the long-distance 
part represents a proton-vacuum matrix element 
of the three quark fields $u(x),u(0),d(0)$.
We  replace this  matrix
element by a  set  of the proton light-cone DAs, equivalent to the nucleon DAs
in the isospin symmetry limit.

Since our main goal  is to 
present the LCSR method and to provide  an approximate 
estimate of the $B\to p$ form factors, we  retain 
only  the lowest twist-3  DAs.
The higher-twist and less singular terms 
of the light-cone expansion in the proton-vacuum matrix element 
are neglected
\footnote{Previous analyses of LCSRs with
the $B$-meson interpolating current revealed that 
higher-twist contributions  are suppressed  by the powers of the inverse Borel mass squared, $M^2\sim m_b\bar{\Lambda}$, where $\bar{\Lambda}$ does not scale with $m_b$.}.

We use the definition of the leading twist-3 DAs from  \cite{Braun:2001tj}: 
\begin{eqnarray}
 &&\bra{0} u_\delta^i(x) u_\beta^j(0) d_\alpha^k(0) \ket{p(P)} = 
\frac{1}{24} \epsilon^{i j k}
\int\limits_0^1 d\alpha_1\alpha_2\alpha_3\,
\delta\Big(1-\sum\limits_{i=1}^{3}\alpha_i\Big)
e^{-i\alpha_1 (P\cdot x)}
\nonumber\\
&&\times\Big[
  \big(\slashed{P}C\big)_{\delta\beta}\big(\gamma_5u_{p}(P)\big)_\alpha
V_1(\alpha_1,\alpha_2,\alpha_3)+
\big(\slashed{P}\gamma_5 C\big)_{\delta\beta}\big(u_{p}(P)\big)_\alpha
A_1(\alpha_1,\alpha_2,\alpha_3)
\nonumber\\
&&+\big(P^\nu i\sigma_{\mu\nu}C\big)_{\delta\beta}\big(\gamma^\mu\gamma_5u_{p}(P)\big)_\alpha
T_1(\alpha_1,\alpha_2,\alpha_3)\Big]\,,
\label{eq:nucleon:DA}
\end{eqnarray}
where the 
variables $\alpha_{1,2,3}$ are the fractions  of the proton
longitudinal momentum carried by the three quarks. 
The DAs $V_1$, $A_1$ and $T_1$ have a polynomial form following from the collinear conformal symmetry.
Adopting the usual next-to-leading approximation in the conformal spin expansion (see e.g. \cite{Braun:2001tj}), 
we use the following expressions:
\ba
&&V_1(\alpha_1,\alpha_2,\alpha_3) = 120f_N\alpha_1\alpha_2\alpha_3\bigg[1+\frac72(1-3V_1^d)
(1-3\alpha_3)\bigg]\,,
\nonumber \\
&&A_1(\alpha_1,\alpha_2,\alpha_3) = 120f_N\alpha_1\alpha_2\alpha_3
\bigg[\frac{21}2A_1^u(\alpha_2-\alpha_1)\bigg]\,,  \nonumber \\
&&T_1(\alpha_1,\alpha_2,\alpha_3) =  120f_N\alpha_1\alpha_2\alpha_3 \bigg[1 - \frac{7}{4}\big(1 - 3 V_1^d - 3 A_1^u\big)\big(1-3\alpha_3\big)\bigg]\,,
\label{eq:DAs}
\ea
where the normalization parameter $f_N$ is the nucleon  decay  constant. 
The two additional parameters $V_1^d$ and $A_1^u$ 
characterize the deviation from the asymptotic form 
of DAs. The latter corresponds 
to $V_1^d=1/3$, $A_1^u=0$.
Below, we will also need the once-integrated DAs:
\ba
&&\widetilde{V}(\alpha_1)\equiv \int_0^{1 - \alpha_1}d\alpha_2\,
\big[V_1(\alpha_1,\alpha_2,1-\alpha_1-\alpha_2)+A_1(\alpha_1,\alpha_2,1-\alpha_1-\alpha_2)\big]
\nonumber\\
&&~~~~~~=20f_N\alpha_1(1-\alpha_1)^3\left(1+
\frac{21}{4}\left[V_1^d-\frac13+A_1^u\right](1-3\alpha_1)\right)\,,
\label{eq:tildV}
\\
&&\widetilde{T}(\alpha_1)\equiv\int_0^{1 - \alpha_1}d\alpha_2\,
T_1(\alpha_1,\alpha_2,1-\alpha_1-\alpha_2) 
\nonumber\\
&&~~~~~~=20f_N\alpha_1(1-\alpha_1)^3\left(1-
\frac{21}{8}\left[V_1^d -\frac13+A_1^u\right](1-3\alpha_1)\right)\;.
\label{eq:tildT}
\ea

\vspace*{0.3cm}
%
Substituting in (\ref{eq:corrq}) the $b$-quark propagator and rewriting the proton-to-vacuum matrix element via DAs 
according to (\ref{eq:nucleon:DA}), we integrate over $x$ and over the virtual 
$b$-quark momentum, obtaining:
\be
\Pi^{(d)}(P,q)= \Bigg[- \frac{ m_b}{2}\int\limits_0^1d\alpha_1
\frac{\big((1-\alpha_1)m_p^2+P\cdot q\big)\widetilde{V}(\alpha_1)}{((1-\alpha_1)P+q)^2-m_b^2}\Bigg]u_{pR}(P)\,.
\label{eq:Pid}
\ee
The expression in square brackets 
is the invariant amplitude $\Pi_R((P+q)^2,q^2)$. We note that
all other amplitudes from the decomposition \eqref{eq:corrdecomp}
vanish in the twist-3 approximation.

To proceed, we transform the invariant amplitude
to a form without $(P+q)^2$ in the numerator:
\ba
\Pi^{(d)}_R((P+q)^2,q^2)=\frac{m_b}{4}\!\int\limits_0^1\!\!
\frac{d\alpha_1 \big(m_b^2+(1\!-\!\alpha_1)^2m_p^2-q^2\big)\widetilde{V}(\alpha_1)}{(1\!-\!\alpha_1)\big[m_b^2+(1\!-\!\alpha_1)\alpha_1 m_p^2-\alpha_1q^2-(1\!-\!\alpha_1)(P+q)^2\big]}\,.
\label{eq:PiR}
\ea
The term independent of $(P+q)^2$ is omitted, because it vanishes after Borel transformation. 
The transition to a dispersion integral form is performed introducing a new integration variable
\be
s=\big[m_b^2-\alpha_1q^2+\alpha_1(1-\alpha_1)m_p^2\big]/(1-\alpha_1)\,,
\label{eq:svar}
\ee
such that the integral $\int_0^1d\alpha_1$ in (\ref{eq:PiR}) transforms to $\int_{m_b^2}^\infty ds$.
Further steps include the quark-hadron duality approximation, 
which replaces the upper limit of the $s$-integration  by 
an effective threshold $s_0^B$ and the Borel transformation 
$1/(s-(P+q)^2)\to e^{-s/M^2}$. After that, we obtain the r.h.s. of (\ref{eq:LCSR}), where
it is convenient to return to the original integration variable $\alpha_1\equiv \alpha$. Our main result, the LCSR 
for the form factor $F^{(d)}_{B\to p_R}(q^2)$, reads:
\be
F^{(d)}_{B\to p_R}(q^2)\!\! \; =
\frac{m_b^3}{4m_B^2f_B}\!\!\int\limits_0^{\alpha_0^B} \!\!d\alpha\, e^{(m_B^2-s(\alpha))/M^2}
\bigg(1+\frac{(1-\alpha_1)^2m_p^2-q^2}{m_b^2}\bigg)
\frac{\widetilde{V}(\alpha)}{(1-\alpha)^2}\,,
\label{eq:LCSRfdR} 
\ee
where $s(\alpha)$ is given by (\ref{eq:svar}) and the effective threshold transforms to 
$$\alpha_0^{B}=\frac{s_0^B-q^2+m_p^2-\sqrt{(s_0^B-q^2+m_p^2)^2-4m_p^2(s_0^B-m_b^2)}}{2m_p^2}\,.$$

Following the same procedure as above, 
we derive the LCSRs for the $B\to p$ form factors in the  model $(b)$. We start, correspondingly,
from the correlation function: 
\ba
&&\Pi^{(b)}_\alpha(P,q)=-i\epsilon^{ijk} m_b\int d^4x\ e^{i(P+q)\cdot x}
\nonumber \\
&&\times\bra{0}T\left\{\big[\bar{b}(x)\gamma_5u(x)\big] ,
b_R^i(0)_\alpha\big[\big(d^{j}_R(0)\big)^TCu_R^k(0)\big]\right\}\ket{p(P)} \,,
\label{eq:corrqb}
\ea
and the OPE result in the twist-3 approximation reads:
\ba
\Pi^{(b)}(P,q)= \Bigg[-\frac{ m_b m_p}{4}\int\limits_0^1d\alpha_1
\frac{(1-\alpha_1)m_p\widetilde{V}(\alpha_1)-3m_b \widetilde{T}(\alpha_1)}{((1-\alpha_1)P+q)^2-m_b^2}\Bigg]u_{pR}(P)
\nonumber
\\
-\Bigg[\frac{ m_bm_p}{4}\int\limits_0^1d\alpha_1\frac{\widetilde{V}(\alpha_1)}{((1-\alpha_1)P+q)^2-m_b^2} 
\Bigg]\slashed{q}u_{pL}(P)\,.
\label{eq:Pib}
\ea
Note that in this case we encounter two different invariant amplitudes, hence two form factors, for which we  obtain
the following LCSRs:

\ba
F^{(b)}_{B\to p_R}(q^2)&=& 
\frac{m_b^2 m_p}{4m_B^2f_B}\!\int\limits_0^{\alpha_0^B} \!\!d\alpha\, e^{(m_B^2-s(\alpha))/M^2}
\bigg(\frac{m_p}{m_b}\widetilde{V}(\alpha)- \frac{3}{1-\alpha}\widetilde{T}(\alpha)\bigg)\,,
\label{eq:LCSRfbR} 
\\
\widetilde{F}^{(b)}_{B\to p_L}(q^2)&=&
\frac{m_b m_p^2}{4m_B^2f_B}\!\int\limits_0^{\alpha_0^B} \!\!d\alpha\, e^{(m_B^2-s(\alpha))/M^2}
\frac{\widetilde{V}(\alpha)}{(1-\alpha)}\,.
\label{eq:LCSRfbL} 
\ea
\begin{table}[h]
\centering
\begin{tabular}{|l|c|c|}
\hline
Parameter  &  interval & [Ref.]  \\[1mm]
\hline
&&\\[-3.0mm]
$b$-quark $\overline{MS}$ mass & $\bar{m}_b(3~\mbox{GeV})= 4.47^{+0.04}_{-0.03} $ GeV 
&\cite{Zyla:2020zbs}\\[1mm]
Renormalization scale &$\mu =3.0^{+1.5}_{-0.5}$ GeV&
\multirow{3}{*}{
\cite{Khodjamirian:2017fxg,Khodjamirian:2020mlb}}
\\[1mm]
Borel parameter squared &  $M^2=16.0\pm 4.0 $ GeV$^2$ &\\[1mm]
Duality threshold  & $ s_0 =39.0^{-1.0}_{+1.5} $ GeV$^2$ &  \\[1mm]
\hline
&&\\[-3.0mm]
$B$-meson decay constant & $f_B= 190.0\pm 1.3$ MeV &\cite{Aoki:2021kgd}\\[1mm]
\hline
&&\\[-3.0mm]
Nucleon decay constant &  $f_N(\mu=2~\mbox{GeV}) = \big(3.54^{+0.06}_{-0.04}\big)\times 10^{-3}~ \mbox{GeV}^2$ 
&
\multirow{3}{*}{\cite{RQCD:2019hps}}
\\[1mm]
Parameter of twist-3 DAs & 
$\varphi_{11}(\mu=2~\mbox{GeV})=0.118^{+0.024}_{-0.023}$&\\[1mm]
\hline
\end{tabular}
\caption{The input parameters in the LCSRs.  }
\label{tab:input}
\end{table}
\vspace{0.3cm}

{\bfseries 6.}~~
The input parameters for LCSRs  are collected in Table~\ref{tab:input}.
The $b$-quark mass  is taken in $\overline{MS}$-scheme.
The adopted renormalization scale and its variation 
are  optimal for a correlation function with the $B$-meson
interpolating current, following e.g. the  recent analyses of  LCSRs for the $B\to\pi$ 
form factors and $B^*B\pi$ strong couplings   \cite{Khodjamirian:2017fxg,Khodjamirian:2020mlb}.
Accordingly, we use the same range for the Borel parameter.  We expect
this range to be large enough for the validity of the leading twist 
approximation in LCSRs, since the yet unaccounted higher twist DAs 
lead to contributions that are typically suppressed
by inverse powers of $M^2$. We also checked that, with the same choice of
the Borel parameter range, the contributions of excited
and continuum states to LCSRs do not exceed 10\% (20\%) for the model $(d)$ (model $(b)$). Hence, our results are not too sensitive to the quark-hadron duality approximation. The effective threshold $s_0^B$ is then fixed for each value of $M^2$, fitting 
to the $B$ meson mass the LCSR differentiated with respect to $-1/M^2$.
Furthermore, for the $B$-meson decay constant we use the average \cite{Aoki:2021kgd} of 
the $N_f=2+1+1$ lattice QCD determinations.

For the input parameters determining the nucleon DAs, we take the most recent lattice QCD results \cite{RQCD:2019hps} at the scale $\mu_0=2.0$ GeV. 
The scale dependence is calculated using the known one-loop 
anomalous dimensions, e.g. for the nucleon decay constant:
\be
f_N(\mu)=\left(\frac{\alpha_s(\mu)}{\alpha_s(\mu_0)}\right)^{2/(3\beta_0)}f_N(\mu_0)\,,
\label{eq:fNrenorm}
\ee
where 
$\beta_0=11-2/3 n_f$ and $n_f=4$ for the scales of our choice.
The parameters $A_1^u$ and $V_1^d$
 are equal to linear combinations of the multiplicatively renormalizable nonasymptotic coefficients $\varphi_{10}$ and $\varphi_{11}$
 (for details see e.g., \cite{Anikin:2013aka}). We notice that the contributions of $\varphi_{10}$ cancel in the  linear 
 combination of $A_1^u$ and $V_1^d$ entering both integrated DAs in (\ref{eq:tildV}) 
 and (\ref{eq:tildT}). As a result, the LCSRs in the leading twist 
 depend on a single nonasymptotic coefficient:
\ba
V_1^d(\mu)-\frac13 +A_1^u(\mu)=
\frac43\Big(\frac{\alpha_s(\mu)}{\alpha_s(\mu_0)}\Big)^{8/(3\beta_0)}\varphi_{11}(\mu_0)\,,
\label{eq:AVrenorm}
\ea
equal to $0.148\pm 0.03$ at $\mu=3 $ GeV, if we use $\varphi_{11}(\mu_0=2~\mbox{GeV})$ from Table~\ref{tab:input}.

With the input specified above,  the $B\to p$ form factors are then calculated from the LCSRs (\ref{eq:LCSRfdR}),
(\ref{eq:LCSRfbR}) and (\ref{eq:LCSRfbL}) at 
$q^2\ll m_b^2$, including also the spacelike region $q^2<0$.
Hence, we can directly obtain the $B\to p\Psi$ amplitude 
at a dark-antibaryon mass, $m_\Psi=\sqrt{q^2}$, in the lower part of the 
expected \cite{Elor:2018twp,Alonso-Alvarez:2021qfd} range $m_p \leq  m_\Psi \leq (m_B-m_p)\simeq 4.34$ GeV.  
To extrapolate the form factors to the upper part of that range, we  
employ the  $z$-expansion, mapping  the complex $q^2$-plane 
onto the unit disk on the plane of the variable
$z(q^2) = (\sqrt{t_{+}-q^2}-\sqrt{t_{+}-t_{0}})/
(\sqrt{t_{+}-q^2}+\sqrt{t_{+}-t_{0}}) $\,, 
where 
$t_{\pm} = (m_B  \pm m_{p})^2$ and
$t_{0} = (m_B + m_p) \cdot (\sqrt{m_B} - \sqrt{m_p})^2\,.$
More specifically, we use the BCL version \cite{Bourrely:2008za} of $z$-expansion, 
slightly modified according to \cite{Khodjamirian:2017fxg}. For instance, we use for the model-$(d)$ 
form factor:
\begin{eqnarray}
F^{(d)}_{B\to p_R}(q^2)= \frac{F^{(d)}_{B\to p_R}(0)}{1 - q^2/m_{\Lambda_b}^2}
\Bigg\{1 + b^{(d)}_{B\to p_R}\Bigg[z(q^2) - z(0)
+ \frac{1}{2} \Big( z(q^2)^2 - z(0)^2\Big)\Bigg] \Bigg\}\,.
\label{eq:BCL}
\end{eqnarray}
Note that the pole factor in (\ref{eq:BCL}) reflects the 
pole of $\Lambda_b$-baryon  located below the $\bar{B}+p$ threshold.
Similar expansions parameterized by the value at $q^2=0$ and by the slope parameter are used 
for the form factors in the model $(b)$. 

For the fit of the form factors from LCSRs to their $z$-expansion, 
we have chosen the interval $-5.0 <q^2<+1.0 $ GeV$^2$ ($0.113\geq z\geq 0.077$). 
The adopted order of this expansion is quite sufficient, since the extrapolation interval  $1.0~\mbox{GeV}^2 <q^2<(m_B-m_p)^2 $ also maps onto small $z$-values, $0.077>z>-0.083$.
The fitted parameters of (\ref{eq:BCL}) and of the analogous expressions for the 
two form factors in the model $(b)$ are presented in Table~\ref{tab:FitParameters}.

\begin{table}[h]
\centering
\begin{tabular}{|l|c||c|c||c|c|}
\hline
&&&&&\\[-2mm]
$F^{(d)}_{B\to p_R}(0)$& $b^{(d)}_{B\to p_R} $ & $F^{(b)}_{B\to p_R}(0)$& $b^{(b)}_{B\to p_R}$& 
$F^{(b)}_{B\to p_L}(0)$& $b^{(b)}_{B\to p_L}$\\[1mm]
\hline
&&&&&\\[-2mm]
$0.026^{+0.002}_{-0.002} $& $ 6.81^{+1.44}_{-0.85} $& $ -0.008_{-0.001}^{+0.002} $&$ -2.08_{-0.29}^{+0.28}$ & $0.0008_{-0.0001}^{+0.0002} $&$ 0.40_{-0.41}^{+0.30} $\\[1mm]
\hline
\end{tabular}
\caption{ Parameters of the $z$-expansion for the $B\to p$ form factors (in GeV$^2$ units).} 
\label{tab:FitParameters}
\end{table}
The quoted uncertainties are obtained
by varying the input parameters within the adopted intervals and
adding them in quadrature.
We observe a significant suppression of the form factors in the model
$(b)$ with respect to the model $(d)$.
This effect can be traced to the $O(m_p^2/m_b^2)$ suppression of the LCSRs
(\ref{eq:LCSRfbR}) and (\ref{eq:LCSRfbL}) 
with respect to (\ref{eq:LCSRfdR}). Without going into further details, we conclude that
there is a nontrivial sensitivity of the hadronic $B\to p$ form factors to the 
configuration of quark fields in the effective operator.

\begin{figure}
\begin{center}
\includegraphics[scale=0.55]{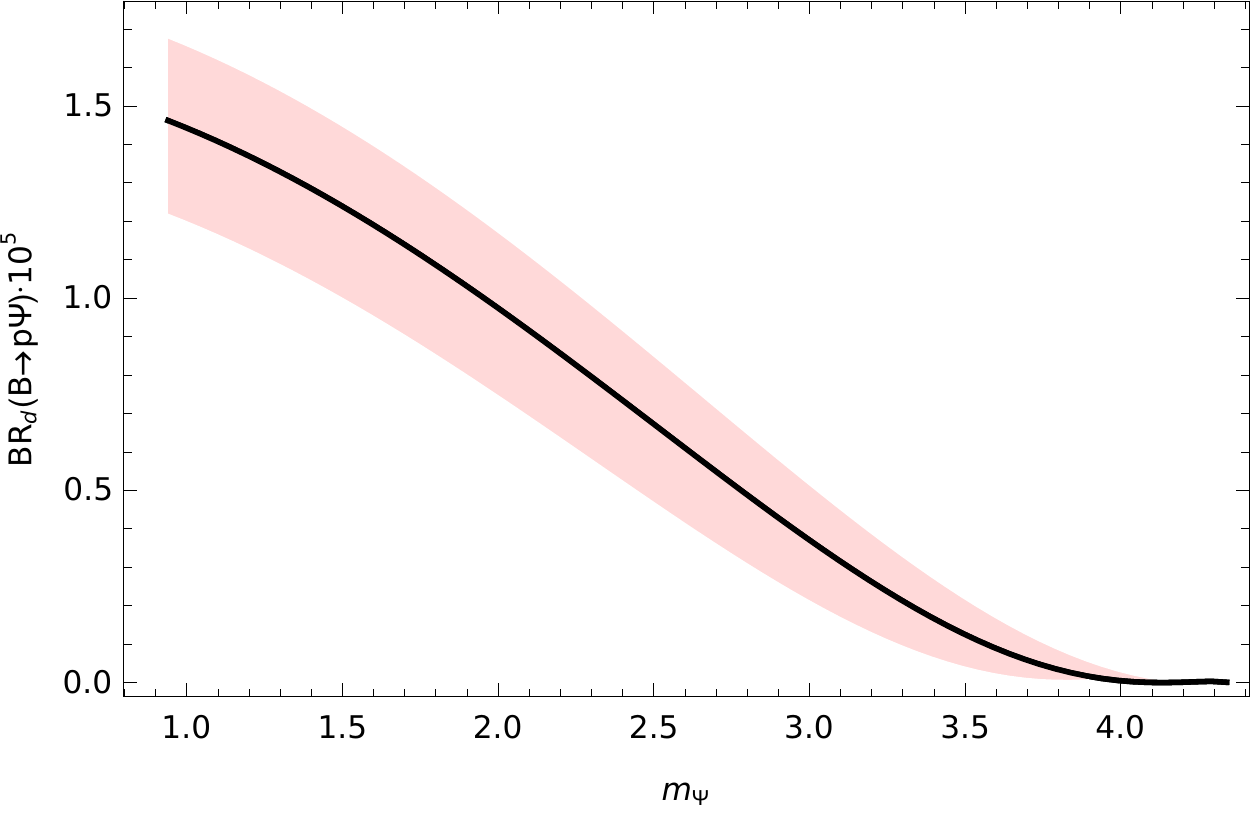}~~~
\includegraphics[scale=0.55]{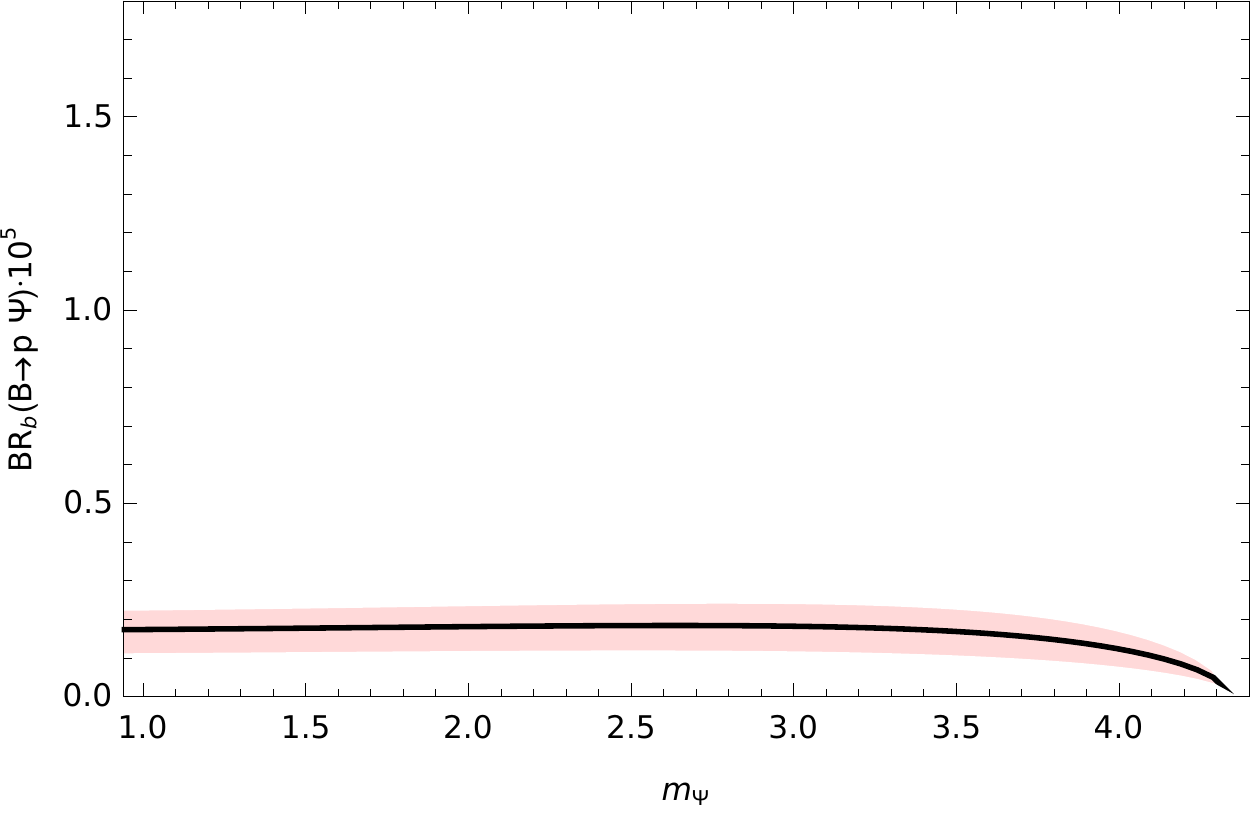}
\end{center}
\caption{Branching fractions of the $B\to p\Psi$ decay as a function of the $\Psi$ mass
at $|G_{(d)}|^2=|G_{(b)}|^2=10^{-13} ~\mbox{GeV}^{-4}$.
The curve (band) on the left and right panels corresponds to  the 
central values of the input (estimated uncertainties) for the model (d) and 
model (b), respectively.}
\label{fig:Branching}
\end{figure}

Finally, substituting 
the form factors in the decay amplitude (\ref{eq:amplAB}) and its analog for the model $(b)$, we 
calculate  the $B\to p \Psi$ decay widths in both models: 
\ba
\Gamma_{(d)}(B^+\to p\Psi)=
|G_{(d)}|^2\Big(F^{(d)}_{B\to p_R}(m_\Psi^2)\Big)^2
\big(m_B^2-m_p^2-m_\Psi^2\big) 
\,\frac{\lambda^{1/2}(m_B^2,m_p^2,m_\Psi^2)}{16\pi m_B^3}\,.
\label{eq:widthd}
\ea
\ba
\Gamma_{(b)}(B^+\to p\Psi)=
|G_{(b)}|^2 \Bigg\{\Bigg[\Big(F^{(b)}_{B\to p_R}(m_\Psi^2)\Big)^2
+ \frac{m_\psi^2}{m_p^2}\Big(\widetilde{F}^{(b)}_{B\to p_L}(m_\Psi^2)\Big)^2\Bigg]
\big(m_B^2-m_p^2-m_\Psi^2\big)
\nonumber\\
+ \;
2 m_\Psi^2F^{(b)}_{B\to p_R}(m_\Psi^2)\widetilde{F}^{(b)}_{B\to p_L}(m_\Psi^2)\Bigg\}
\,\frac{\lambda^{1/2}(m_B^2,m_p^2,m_\Psi^2)}{16\pi m_B^3}\,,
\label{eq:widthb}
\ea
where in the last expression we take into account that
the form factors are real valued in the region below $(m_B+m_p)^2$. 
Multiplying  the above expressions by the $B^\pm$ lifetime, $\tau_{B^\pm}=1.638 \pm 0.004$ ps 
\cite{Zyla:2020zbs},
we arrive at our main results, the branching fractions of the $B^+\to p\Psi$ decay. They are plotted in Fig.~\ref{fig:Branching}. For uniformity, we assume equal effective four-fermion couplings $|G_{(d)}|^2=|G_{(b)}|^2=10^{-13}~\mbox{GeV}^{-4}$.
This choice is in the ballpark of the upper limits extracted \cite{Alonso-Alvarez:2021qfd} 
from the LHC constraints on the heavy coloured scalars $Y$ predicted in the B-Mesogenesis models. 
Our predictions for the branching fractions can be easily rescaled for any other value
of these couplings. \\

{\bfseries 7.}~~
In this letter, we presented the first estimate of the  
$B$-meson decay width into a proton and dark antibaryon $\Psi$ emerging  
in the $B$-Mesogenesis framework. Our main results are the $B\to p $ form factors determining the hadronic amplitude of this decay. 
They were calculated as a function of $\Psi$ mass from the QCD-based LCSRs with the proton distribution amplitudes.
Our results reveal 
a strong dependence of these form factors on the form of the
effective operator originating 
from the three-quark coupling with a dark antibaryon.

The accuracy of our calculation is limited by the 
lowest twist of the nucleon DAs and  by the leading $O(\alpha_s^0)$ of the underlying OPE. Both higher-twist and $O(\alpha_s)$ contributions
can be systematically calculated in the future. However, we do not expect a significant impact of these contributions
on the numerical results obtained here.

It is straightforward to apply the LCSR method suggested here 
to other versions of the $B$-Mesogenesis scenario, e.g. to a superposition of the models $(d)$ and $(b)$,
or to the models \cite{Alonso-Alvarez:2021qfd} with the hypercharge $Q_Y=+2/3$ of the heavy mediator. Other exclusive $B$-decays, for example,
into a strange baryon and $\Psi$, are also accessible with a 
certain modification of the LCSRs. A complete assessment  of the $B\to \mbox{hadrons} +\Psi$ decays is indispensable without reliable estimates
of their inclusive widths. Here, one has to employ quite different methods, e.g.
the well established heavy-quark expansion.

Finally, our results can be useful for ongoing and future experimental searches 
for the $B$-decays to invisibles.  According to \cite{Alonso-Alvarez:2021qfd}, 
the Belle II experiment is sensitive to  branching fractions in the ballpark of  $BR(B\to p \Psi)=3\times 10^{-6}$. Comparing with our results plotted in Fig.~\ref{fig:Branching}, we conclude that, at least for the  model
$(d)$, such a sensitivity  is well within the parameter
space formed by the limits on the effective coupling $G_{(d)}$ and the mass $m_\Psi$, 
 provided this mass is fixed by measuring the proton energy in the $B$ meson rest frame.
This enables a realistic test of the $B$-Mesogenesis scenario in the future.

\subsection*{Acknowledgments}

This work was supported by the Deutsche Forschungsgemeinschaft 
(DFG, German Research Foundation) under the grant 396021762 - TRR 257. 
A.K. acknowledges the hospitality of the Munich Institute for Astro\,- and Particle Physics (MIAPP) 
which is funded by the DFG
under Germany's Excellence Strategy – EXC-2094 – 390783311 and 
where the  part of this work  has been done.


\end{document}